\documentclass[preprint]{aastex631}

\usepackage{soul} 

\shorttitle{Winds from starburst rings}
\shortauthors{Nguyen \& Thompson}
\graphicspath{{./}{figures/}}

\usepackage{amsmath}

\begin{document}

\title{Galactic winds and bubbles from nuclear starburst rings}

\author[0000-0002-1875-6522]{Dustin D. Nguyen}
\thanks{2\textit{d} and 3\textit{d} movies of the simulations can be found online at \href{ https://dustindnguyen.com/simulation_pages/nguyen_thompson_2022_rings/}{dustindnguyen.com}.}
\affiliation{Center for Cosmology and Astro-Particle Physics (CCAPP) \\
The Ohio State University, Columbus, Ohio, 43210, USA}

\affiliation{Department of Physics, The Ohio State University, Columbus, Ohio, 43210, USA}
\author[0000-0003-2377-9574]{Todd A. Thompson}
\affiliation{Center for Cosmology and Astro-Particle Physics (CCAPP) \\
The Ohio State University, Columbus, Ohio, 43210, USA}
\affiliation{Department of Physics, The Ohio State University, Columbus, Ohio, 43210, USA}
\affiliation{Department of Astronomy, The Ohio State University, Columbus, Ohio, 43210, USA}



\begin{abstract}


Galactic outflows from local starburst galaxies typically exhibit a layered geometry, with cool $10^4\,$K flow sheathing a hotter $10^7\,$K, cylindrically-collimated, X-ray emitting plasma. Here, we argue that winds driven by energy-injection in a ring-like geometry can produce this distinctive large-scale multi-phase morphology. The ring configuration is motivated by the observation that massive young star clusters are often distributed in a ring at the host galaxy's inner Lindblad resonance, where larger-scale spiral arm structure terminates. We present parameterized three-dimensional radiative hydrodynamical simulations that follow the emergence and dynamics of energy-driven hot winds from starburst rings. In this Letter, we show that the flow shocks on itself within the inner ring hole, maintaining high $10^7$\,K temperatures, whilst flows that emerge from the wind-driving ring unobstructed can undergo rapid bulk cooling down to $10^4\,$K, producing a fast hot bi-conical outflow enclosed by a sheath of cooler nearly co-moving material without ram-pressure acceleration. The hot flow is collimated along the ring axis, even in the absence of pressure confinement from a  galactic disk or magnetic fields. In the early stages of expansion, the emerging wind forms a bubble-like shape reminiscent of the Milky Way's eROSITA and Fermi bubbles and can reach velocities usually associated with AGN-driven winds. We discuss the physics of the ring configuration, the conditions for radiative bulk cooling, and the implications for future X-ray observations. 

\end{abstract}

\keywords{\href{http://astrothesaurus.org/uat/17}{Active galaxies (17)}; \href{http://astrothesaurus.org/uat/1570}{Starburst galaxies (1570)}; \href{http://astrothesaurus.org/uat/929}{Galaxy evolution (929)}; \href{http://astrothesaurus.org/uat/767}{Hydrodynamic simulations (767)}; \href{http://astrothesaurus.org/uat/1602}{Stellar feedback (1602)}; \href{http://astrothesaurus.org/uat/1602}{Astronomical simulations (1857)}; \href{http://astrothesaurus.org/uat/1602}{Galactic winds (572)} }


\section{Introduction} \label{sec:intro}

Galactic winds are important to galaxy evolution. They drive gas into the circumgalactic medium, help truncate star formation, and set the mass-metallicity relation \citep[see reviews by][]{Veilleux2005,Zhang2018}. Energy-driven galactic winds, powered by massive star supernovae (SNe) and stellar winds, may generate the hot ($T\approx 10^6-10^7\,\mathrm{K}$) gas seen in diffuse X-ray emission from local starbursts  \citep{Strickland2004,Strickland2007}. While the hot gas kinematics have not yet been directly measured, current models predict velocities of $\sim500-2000\,\mathrm{km\,s^{-1}}$, depending on the mass-loading of the hot phase \citep{Chevalier1985,Strickland2009,Nguyen2021}.
In the M82 superwind, other local starbursts, and high-redshift rapidly star-forming galaxies, the cooler neutral and ionized gas is observed to be moving at velocities of order $\simeq 500-1000\,\mathrm{km\,s^{-1}}$ 
\citep{Shopbell1998,Shapiro2009}. A motivating fact for this paper is that in local systems where the different thermal components can be spatially resolved, the hot and cool outflows exhibit a layered geometry, with the more cylindrical hot gas surrounded by a cooler gas ``sheath" of confining medium (e.g., as seen in M82, NGC 253, NGC 3079, and others; \citealt{Westmoquette2009,Leroy2015,Hodges-Kluck2020}). 

A critical open issue in galactic wind physics is the acceleration mechanism for the cool component. One well-known picture is that the cool phase is composed of clouds outside of the star-forming core that are ram pressure accelerated by the emergent hot supersonic wind. Simulations of this interaction often lead to the destruction of clouds by hydrodynamical instabilities \citep{Cooper2009,Scannapieco2015,Schneider2018}. In some regimes of cloud parameters, the wind-cloud interaction leads to cloud growth and acceleration \citep{Gronke2018}, although this process has not yet been shown to operate in global galactic wind simulations \citep{Cooper2007,Tanner2016,Schneider2020}. Another possible solution is that the cool phase condenses directly from the hot phase via radiative instability \citep{Wang1995,Silich2004,Tenorio-Tagle2003,Thompson2016}. The condition for rapid cooling is set primarily by the physical size of the wind-driving region and the hot wind mass-loading rate. 

Here, we argue that the characteristic layered outflow geometry seen in nearby starbursts may be produced by galactic winds driven from a ring-like geometry. Rings of massive star clusters form as a result of gas inflow to the inner Lindblad resonance  \citep{Athanassoula1992,Binney2008}. These rings often dominate the total star formation rate within their host galaxies \citep{Mazzuca2008}. For example, the circum-nuclear ring of the archetypal nearby starburst galaxy M82 is well studied. \citet{Nakai1987} discovered a double-lobed distribution of CO, which is interpreted as a molecular torus. \citet{Hughes1994} studied thermal reradiated emission from dust in sub\,mm wavelengths and found that the intensity profile can be modeled as an edge-on nuclear starburst torus. \cite{Weiss2001} provided a multi-transition analysis of the molecular gas and showed the regions of most violent star formation are confined to the molecular lobes, and are arranged in a toroidal topology around the nucleus. Star-forming rings are also observed in the nearby starburst NGC 253 \citep{Arnaboldi1995,Leroy2018,Levy2022},  other local galaxies \citep[see][]{Buta1993,Boker2008,Comer2010,Leroy2021}, and the Milky Way galaxy \citep[see][]{Henshaw2022}. 
3\textit{d} ring simulations have focused on the dynamical processes associated with stellar populations and bars, and are typically carried out with parameters characteristic of the Galaxy's Central Molecular Zone \citep{Armilotta2019,Tress2020}. \citet{Moon2021a} found that individual SN shocks can merge together to create large hot outflows resembling galactic winds (see also \citet{Tenorio-Tagle2003}). These works motivate a focused study of the general properties, dynamics, and morphology of energy-driven winds from star-forming rings. Here, we posit the ring structure and focus on the evolution of large-scale outflows that can be produced from parameterized energy and mass injection within the ring volume. As nuclear rings are observed to substantially vary in size, ranging from diameters of tens of pc to several kpc \citep{Buta1993,Comer2010}, we consider several different ring sizes and mass-loading rates. In addition, we consider two ring geometries: the first is an ideal azimuthally uniform ring, which is used to explore the basic structures that can be obtained from this configuration. The second is a series of uniform spheres distributed in a ring-like geometry that is meant to mimic individual super star clusters formed at the inner Lindblad resonance \citep{Elmegreen1994}. For both sets of simulations, we find that a ring-like distribution of sources can produce an overall wind morphology akin to that seen in observations, where the hot and cool phases are layered, and where the hot flow is confined to a more cylindrical structure. 

\begin{figure*}[h!]
    \centering
    \includegraphics[width=\textwidth]{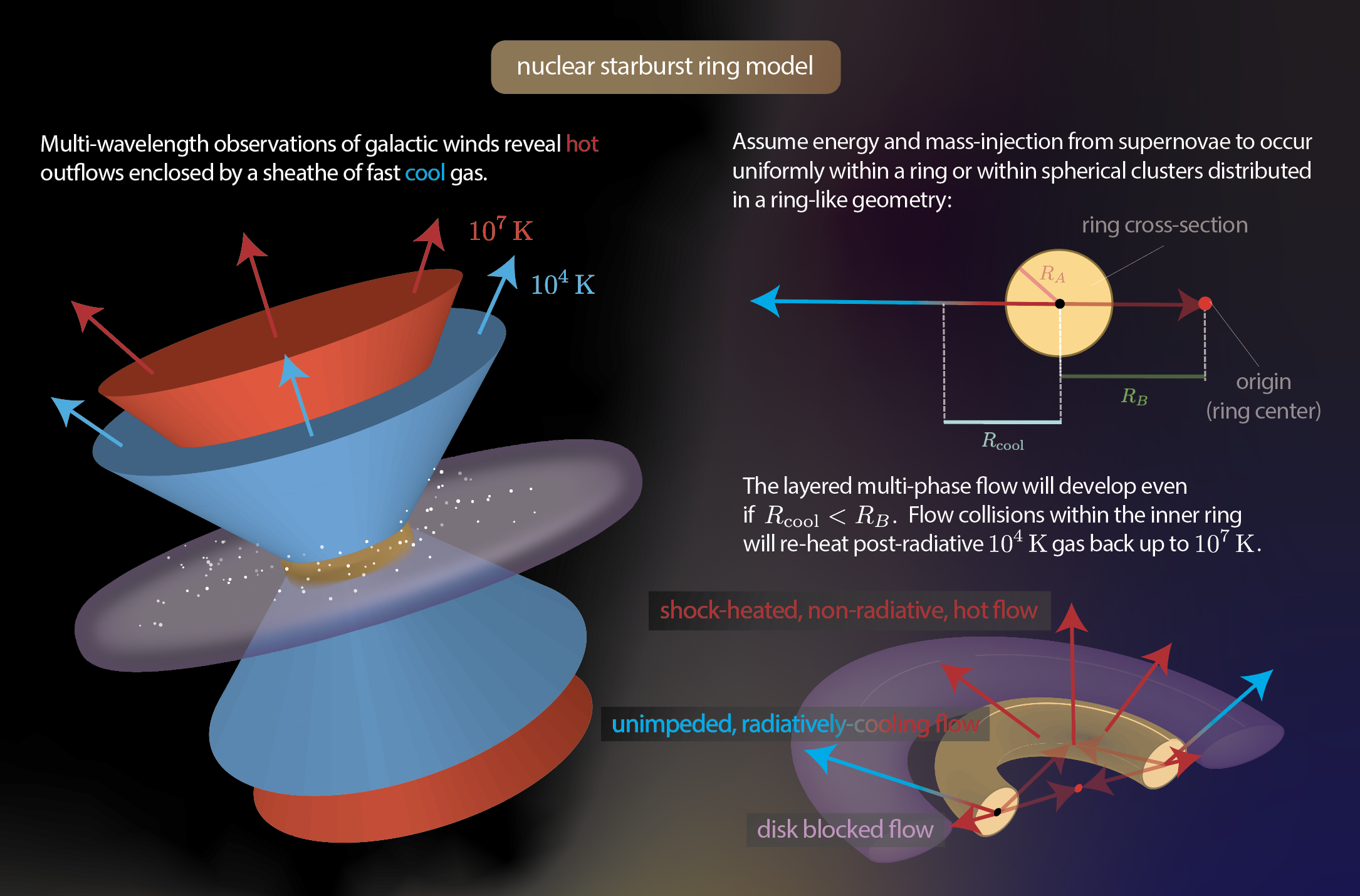}
    \caption{Schematic of the nuclear starburst ring model. In this model, supernovae inject energy and mass into a ring geometry, leading to a steady state flow that has an initial temperature within the ring of $\sim 10^7\,$K, which accelerates to high velocity as it leaves the ring volume. For high mass-loading, the wind emanating from the outer-ring can become radiative and quickly cool to $10^4\,$K at the cooling radius, $r_\mathrm{cool}$ (see Section \ref{sec:results}). Within the ring hole, along the minor axis, the flow shocks on itself, thermalizing its kinetic energy and maintaining high temperature. As the hot flow emerges from the inner ring region, it is collimated by the surrounding cooler flow.}
    \label{fig:donut}
\end{figure*}

\newpage 

\section{Model and Methods} \label{sec:methods}
In Fig.~\ref{fig:donut} we illustrate the model problem. We assume SNe deposit energy and mass at total rates $\dot{E}_\mathrm{T}$ and $\dot{M}_\mathrm{T}$, uniformly throughout a symmetric ring with volume $V$ (see Appendix~\ref{sec:EdotMdot}). The controlling parameters of the problem are the source terms, the energy and mass-loading rates, the ring width $R_A$, and the ring radius $R_B$. These quantities are shown in Table \ref{tab:sims} and are chosen for Simulations `\texttt{A}' and `$\texttt{A}_\star$' such that the unobstructed flow emanating from the outer ring undergoes bulk cooling. 

We use the GPU accelerated radiative hydrodynamics code \texttt{Cholla} \citep{Schneider2015} to run a suite of simulations to study the properties of the starburst ring model. The simulations and parameters are summarized in Table \ref{tab:sims}. The simulation box is $10 \times 10 \times 15\,$kpc for the $\hat{x} \times \hat{y} \times \hat{z}$ directions, respectively. There are 2 high resolution simulations, `$\texttt{A}$' and `$\texttt{A}_\star$', with $768 \times 768 \times 1152$\,cells in each spatial dimension, giving a cell resolution of $\Delta x \simeq 13$\,pc. These two serve as the fiducial simulations for the study. The uniform ring in Simulation `\texttt{A}' (see Section \ref{sec:uniform_ring}) has a ring width of $R_A=25\,$pc. Simulation `$\texttt{A}_\star$' (see Section \ref{sec:nonuniform_ring}) has injection regions modeled as 8 identical spherical star clusters distributed in a ring. Each star cluster has a radius of $R_A=25\,$pc. The other 10 runs, Simulations `\texttt{B}...' and `\texttt{C}...', are run with $512 \times 512 \times 768$\,cells, giving a cell resolution of $\Delta x \simeq 19.5$\,pc. These lower resolution simulations explore the effects to the flow when the ring geometry, mass-loading rate, or ambient circumgalactic gas density is varied. Within the starburst volume, we inject energy and mass at rates $ \dot{E}_\mathrm{T} = \alpha \times 3.1 \times 10^{41} \times  (\dot{M}_\mathrm{SFR}/\mathrm{M_\odot \ yr^{-1}})\,[\mathrm{ergs \ s^{-1}}]$ and $\dot{M}_\mathrm{T} = \beta \times \dot{M}_\mathrm{SFR}$, where $\alpha$, $\beta$, and $\dot{M}_\mathrm{SFR}$ are the thermalization efficiency, mass-loading rate, and star formation rate, respectively. The cut-off in the mass and energy injection rates (see equations \ref{eq:qdot} and \ref{eq:Qdot}) at the ring edge yields a characteristic solution similar to \citet{Chevalier1985} and \citet{Wang1995}, such that a sonic point forms at the edge of the ring volume. 

The Cholla simulations \citep{Schneider2015} use a piece-wise linear interface reconstruction, an HHLC Riemann solver, and an unsplit Van Leer integrator, and a cooling curve that is a piece-wise parabolic fit to a Cloudy curve assuming collisional ionization equilibrium for a solar metallicity gas \citep[see][]{Schneider2018CGols}. We approximately account for background UV heating through the implementation of a cooling floor at $T_\mathrm{floor} = 10^4\,$K. The cooling everywhere is assumed to be optically-thin. Because we do not include a disk in these simulations, the wind from the ring escapes unobstructed, even along the azimuthal plane. These flows would be blocked by the gas disk (see Fig.~\ref{fig:donut}) in a more realistic simulation setup, which we save for future work. Simulations `\texttt{A}', `$\texttt{A}_\star$', `\texttt{B}...' begin with a constant static background of temperature $T_\mathrm{bkgd.}=10^6\,$K and density $n_\mathrm{bkgd.}=10^{-5}\,\mathrm{cm^{-3}}$. Simulations `\texttt{C}...' also start with a static background $10^6\,$K, however the background density is varied for each simulation (see Table \ref{tab:sims}). At the start of the simulation, energy and mass is instantaneously deposited into the uniform (`\texttt{A}'), or distributed cluster (`$\texttt{A}_\star$'), ring geometry. 


Simulations `\texttt{A}' and `$\texttt{A}_\star$' produce radiative flows. As we discuss in Section \ref{sec:results}, the condition for bulk cooling is set primarily by the mass-loading term $\beta$ and the physical width of the wind-driving region $R_A$, as in the analogous spherical models \citep{Thompson2016}). Models with lower mass-loading and/or thicker ring widths (see Appendices \ref{sec:betas} and \ref{sec:beta_trans}) do not produce radiative flows, and a hot wind emerges in all directions.

\begin{table}
    \centering
    \hspace*{-1.5cm}\begin{tabular}{c c c c c c c}
    \hline
    \hline
    Simulation Name & Ring Type & $\alpha$ & $\beta$ & $R_A\,$[pc] & $R_B\,$[kpc] & $\Delta x\,\mathrm{[pc]}$\\
    \hline
    \texttt{A}  \hspace*{0.25cm} & uniform & 1.0 & 0.6 & 25 & 1.0     & 13.0 \\
    \texttt{A}$_\star$  \hspace*{0.09cm} & non-uniform & 1.0 & 0.6 & 25 & 1.0 & 13.0 \\
    \texttt{B0} \hspace*{0.04cm} & uniform & 1.0 & 0.6 & 100 & 1.0     & 19.5 \\
    \texttt{B1} \hspace*{0.04cm} & uniform & 1.0 & 0.6 & 100 & 0.5     & 19.5 \\
    \texttt{B2} \hspace*{0.04cm} & uniform & 1.0 & 0.6 & 100 & 2.0     & 19.5 \\
    \texttt{B3} \hspace*{0.04cm} & uniform & 1.0 & 0.6 & 50 & 1.0     & 19.5 \\
    \texttt{B4} \hspace*{0.04cm} & uniform & 1.0 & 0.6 & 200 & 1.0     & 19.5 \\
    \texttt{B5} \hspace*{0.04cm} & uniform & 1.0 & 1.1 & 100 & 1.0     & 19.5 \\
    \texttt{B6} \hspace*{0.04cm} & uniform & 1.0 & 0.2 & 100 & 1.0     & 19.5 \\
    \texttt{C0} \hspace*{0.04cm} & uniform & 1.0 & 0.1 & 25 & 1.0     & 19.5 \\
    \texttt{C1} \hspace*{0.04cm} & uniform & 1.0 & 0.1 & 25 & 1.0     & 19.5 \\
    \texttt{C2} \hspace*{0.04cm} & uniform & 1.0 & 0.1 & 25 & 1.0     & 19.5 \\
    \hline
    \hline
    \end{tabular}
    \caption{Summary of the simulation parameters. Simulations `\texttt{A}' and `$\texttt{A}_\star$' serve as the fiducial models of the study. Simulations `\texttt{A}', `$\texttt{A}_\star$', and `\texttt{B}...' explore the dynamics and morphology of the steady-state wind, when the initialized gas background at $t=0$ is already blown out of the simulation domain (see Appendices \ref{sec:donut_geo} and \ref{sec:betas}). Simulations `\texttt{B}...' explore steady-state wind properties for values of the mass-loading parameter $\beta$. Simulations `\texttt{C0}', `\texttt{C1}', and `\texttt{C2}' explore the early interaction between the flow and the initialized background gas, with varied background densities of $n_\mathrm{bkgd.}=10^{-3},\ 10^{-5}, \ \mathrm{and} \ 10^{-7}\,\mathrm{cm^{-3}}$, respectively (see Appendix \ref{sec:beta_trans}). The dimensionless parameters $\alpha$ and $\beta$ are characterize the total energy and mass-injection rates, $\dot{E}_\mathrm{T} = \alpha \times 3.1 \times 10^{41} \times (\dot{M}_\mathrm{SFR}/M_\odot \, \mathrm{yr^{-1}) \, [ergs \, s^{-1}]}$ and $\dot{M}_\mathrm{T} = \beta \times \dot{M}_\mathrm{SFR}$, respectively, where $\dot{M}_\mathrm{SFR} = 10\,\mathrm{M_\odot \, yr^{-1}}$.}
    \label{tab:sims}
\end{table}

\newpage 

\section{Results} \label{sec:results}
\subsection{Uniform Ring} \label{sec:uniform_ring} 
\begin{figure}[h!]
    \centering
    \includegraphics[width=\textwidth]{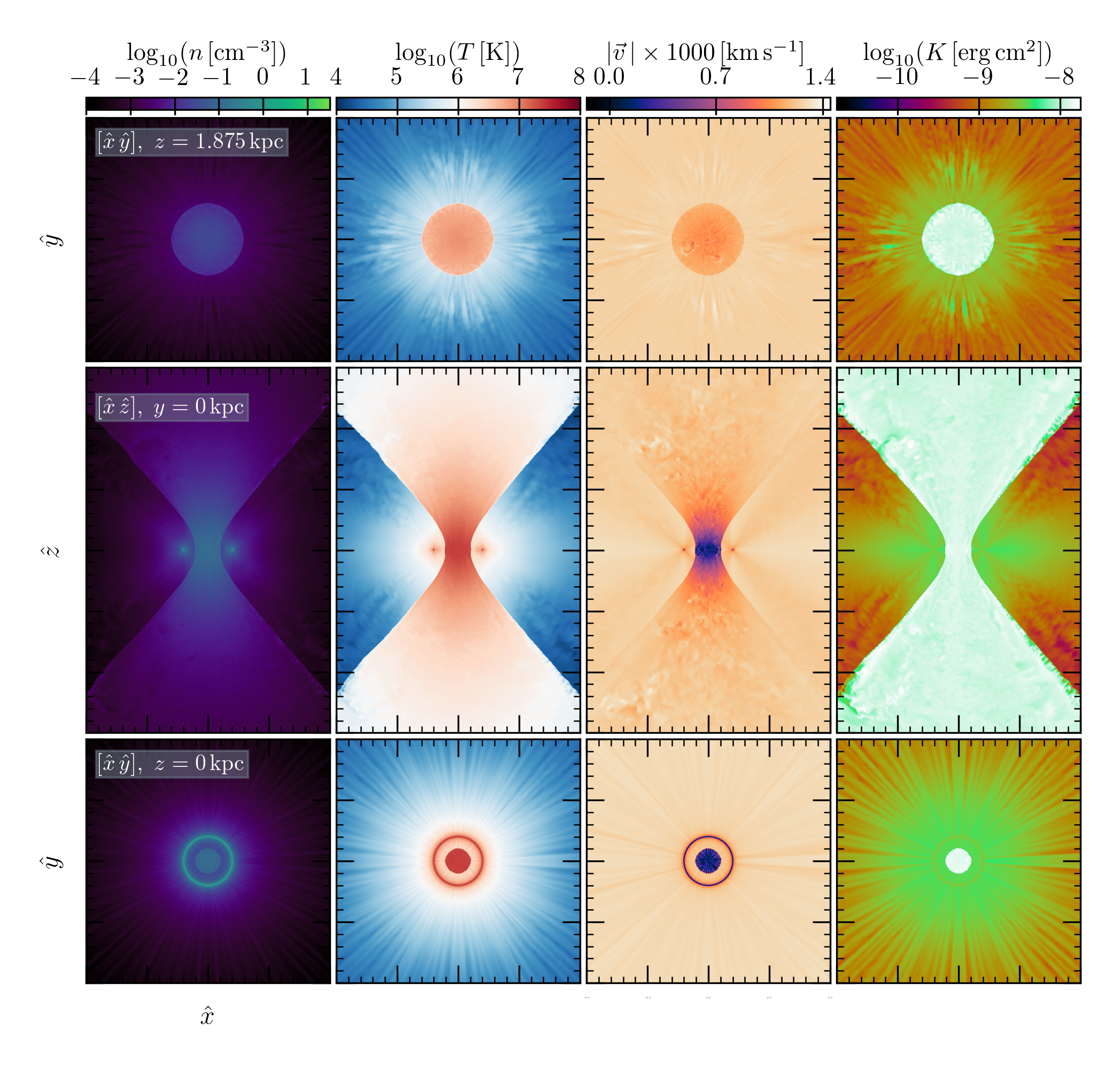}
    \caption{2\textit{d} slices of density, temperature, velocity, and entropy at 20\,Myr from Simulation `\texttt{B0}'. The top row shows the $\hat{x}\hat{y}$ plane slice at $z=1.875\,$kpc, the middle row shows the $\hat{x}\hat{z}$ plane slice at $y=0\,$kpc, and the bottom row shows the $\hat{x}\hat{y}$ plane at $z=0\,$kpc. The starburst ring (bottom row) produces a multi-phase outflow (middle row) where the hot phase is enveloped with a cooler phase (top row). The starburst ring model quickly reaches a steady-state within the first few Myr after the initial transient passes. The simulation size is $10 \times 10 \times 15\,\mathrm{kpc}$ ($\hat{x}\times \hat{y} \times \hat{z}$). Both the top and bottom rows are $10 \times 10$\,kpc and the middle row is $10 \times 15$\,kpc.}
    \label{fig:3views_plainb}
\end{figure}

We first discuss the results for the uniform ring model from Simulation `$\texttt{A}$' (see Table \ref{tab:sims}). In Fig.~\ref{fig:3views_plainb}, we plot three different 2\textit{d} slices of the gas number density, temperature, velocity, and entropy at 20\,Myr of Simulation `\texttt{A}'. Each row displays a different 2\textit{d} slice through the volume. The bottom row is a slice through the origin in the $\hat{x}\hat{y}$-plane, which is a slice through the ring plane. The middle row is a slice through the $\hat{x}\hat{z}$-plane, showcasing the bi-conical hot outflow. The top row is another 2\textit{d} slice through the $\hat{x}\hat{y}$-plane, but at a height of $z=1.875\,$kpc. As seen middle row panels of Fig.~\ref{fig:3views_plainb}, the hot unimpeded flow emerging away from the ring undergoes bulk cooling, while the inner wind material shocks, producing a hot bi-conical flow with a flared-cylinder geometry. These effects are apparent in 1\textit{d} skewers of the unimpeded and shocked flows, shown below. After an initial transient (discussed more and shown in Fig.~\ref{fig:transient}), the system quickly reaches a steady-state within the first few Myr. Because the resolution is comparable to the ring width, there are grid effects. 

\begin{figure}[h!]
    \centering
    \includegraphics[width=\textwidth]{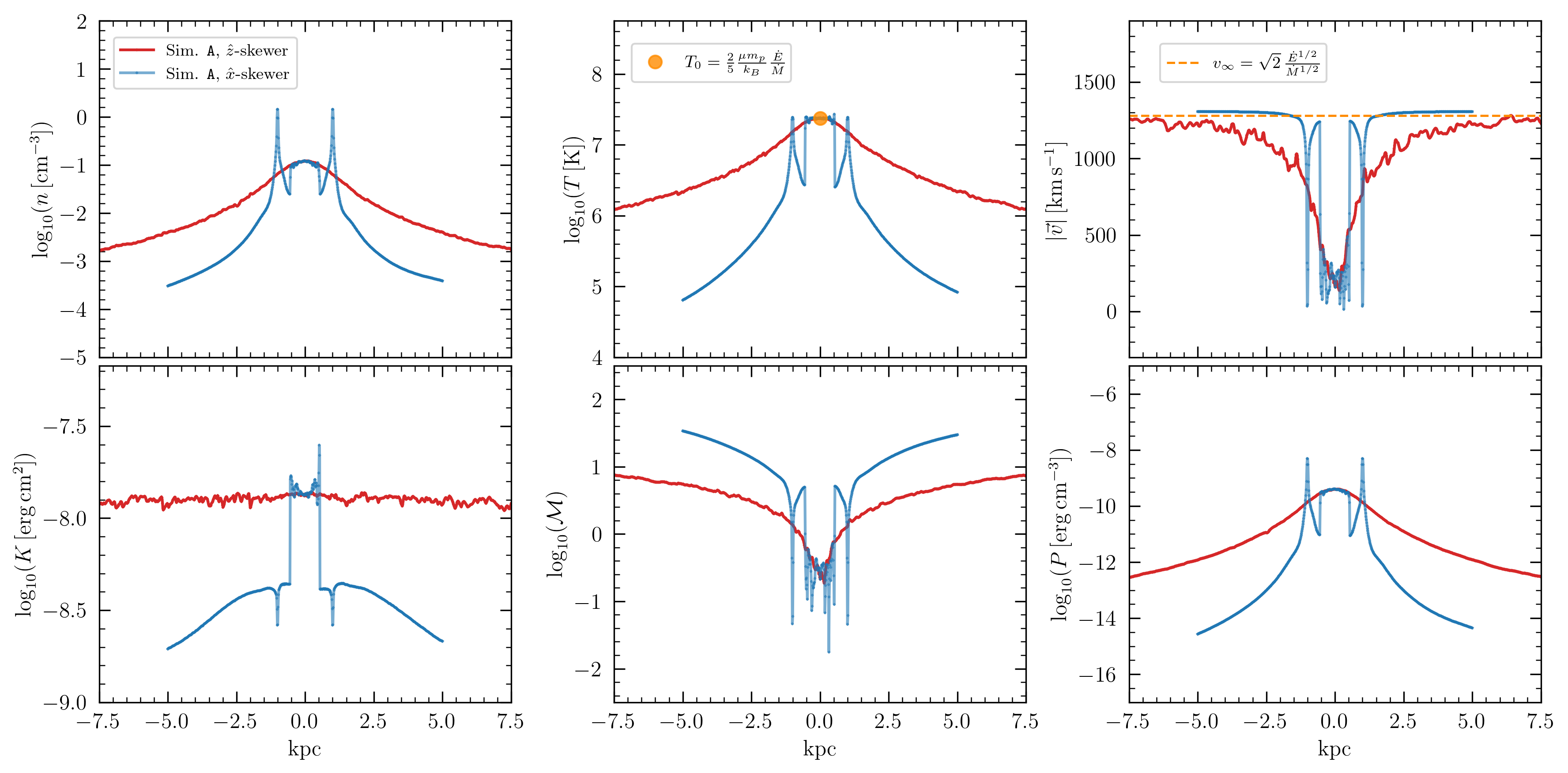}
    \caption{1\textit{d} $\hat{z}$-skewer (red lined-markers), 1\textit{d} $\hat{x}$-skewer (blue lined-markers), and the analytic solutions to the shocked gas temperature and asymptotic velocity in orange. The $\hat{z}$ and $\hat{x}$ skewers pass through the origin at $x=y=0$ and $y=z=0$, respectively. The $\hat{z}$ skewer goes along the minor axis, in the direction of the hot collimated wind.}
    \label{fig:skew}
\end{figure}

Fig.~\ref{fig:skew} shows number density, temperature, velocity, entropy, Mach number, and pressure profiles for 1\textit{d} $\hat{z}$ (red) and $\hat{x}$ (blue) skewers.  The $\hat{z}$ and $\hat{x}$ skewers pass through the origin at $x=y=0$ and $y=z=0$, respectively. The $\hat{z}$ skewer goes along the minor axis, where the flow is collimated. For this $\hat{x}$-skewer (blue line) the peaks, most notably in the upper left and middle panels, correspond to the region where energy and mass is injected into the ring. Looking at the blue line in the upper middle panel and starting at one of the peaks, we see that the flow begins to under-go cooling as it moves towards the origin and is then abruptly shock heated back to $10^7\,$K. This compresses the flow, increasing the density and pressure, while thermalizing its kinetic energy, pushing the flow back into the subsonic $\mathcal{M}<1$ regime (see middle lower panel, $\hat{x}$-skewer, blue line). The effects of radiative cooling are most apparent in the entropy profile (bottom left panel). The hot collimated flow along the minor axis ($\hat{z}$-skewer, red line) maintains a constant entropy. Conversely, the unobstructed flow ($\hat{x}$-skewer, blue line) undergoes strong cooling with a rapid decrease in entropy as a function of distance. The distance at which the flow becomes marginally radiative is most apparent in the entropy profile (bottom left panel of Fig.~\ref{fig:skew}), where the entropy gradually decreases at around $\pm 2.5\,$kpc, or about 1.75 kpc away from the ring. Near the base of the outflow, the hot collimated flow along the minor axis (red line) has shallower temperature, density, and velocity gradients than a spherical flow with the same parameters, as expected from \cite{Nguyen2021}. The asymptotic velocities of both the collimated hot flow and the unimpeded cool flow are: $v_\infty = \sqrt{2} \, \dot{E}_T^{1/2} \dot{M}_T^{-1/2} \simeq 992 \, (\alpha / \beta )^{1/2} \, \mathrm{km \ s^{-1}} \simeq 1280\,\mathrm{km\,s^{-1}}$. The temperature of the shocked gas reaches equivalent temperatures of the uniform wind-driving ring (see the central peaks, blue line in Fig.~\ref{fig:skew}). We find the central temperature of the ring-hole to be identical to that derived by \citet{Chevalier1985} for a spherical wind-driving region (hereafter CC85).
In CC85, $\dot{E}_T$ and $\dot{M}_T$, fill a sphere, whereas here, it is the supersonic wind, which has already left the wind-driving ring, feeding into the shocked volume that supplies the energy and mass. In the center of the ring hole, the kinetic energy is thermalized, such that (see \citet{Strickland2009}) $T_0 = T_\mathrm{ring} = 0.4 \, \dot{E}_T/\dot{M}_T \, \mu m_p/k_B  = 1.43 \times 10^7 (\alpha / \beta) \,\mathrm{K} =2.38 \times 10^7\,K $, where $\mu m_p$ and $k_B$ are the average particle mass and Boltzmann constant, respectively. 

The hot wind is non-spherical. We calculated transverse (with respect to the minor axis) areas for each height $z$ for Simulation `\texttt{A}' and a functional fit  to these values. We model the flow as a flared-cylinder \citep{Breitschwerdt1991,Nguyen2021} with $A(z) = A_0 [1+(z/z_b)^\Delta$]. The flared-cylinder function maintains an approximately constant area $A_0$ up until height $z_b$ when it transitions to a constant solid angle power-law expansion ($\Delta=2$ for spherical flow).  The best fit of the flared-cylinder has $A_0=0.54\,\mathrm{kpc^2}$, $z_b=0.58\,$kpc, and $\Delta=2.14$. The term $d \ln A /dz$ describes the areal divergence rate of the flow (e.g., $A=\Omega r^2$ and $d \ln A / dr = 2/r$ for constant $\Omega$). The flared-cylinder has a slower areal divergence rate than a spherical flow for $z \lesssim 1.5\,$kpc, corresponding to the flow through the cylinder of $A(z)$. However, it eventually matches closely with the spherical expansion model, implying the flared-cylinder fit approaches a constant solid angle expansion rate.  Slower-than-spherical areal flows lead to shallower temperature and density gradients \citep{Nguyen2021}. The combination of the kinetic energy thermalization and cylindrical flow geometry mean that the emerging flow does not become radiative along the minor axis. This is in contrast to the unobstructed flow, which becomes radiative (see entropy profiles, Figs.~\ref{fig:3views_plainb} and \ref{fig:skew}). This occurs because the cooling rate is dependent on the gas temperature, and at the sustained temperatures of $T \geq 10^6$, is at least an order of magnitude less than that of the unimpeded flow when it is at $10^5\,$K. The inferred hot cylindrical flow geometries, described by the $A(z)$ fit, are different for both varied mass-loading rates $\beta$ (see Appendix \ref{sec:betas}) and varied ring-geometries (see Appendix \ref{sec:donut_geo}). 

\begin{figure}[h!]
    \centering
    \includegraphics[width=\textwidth]{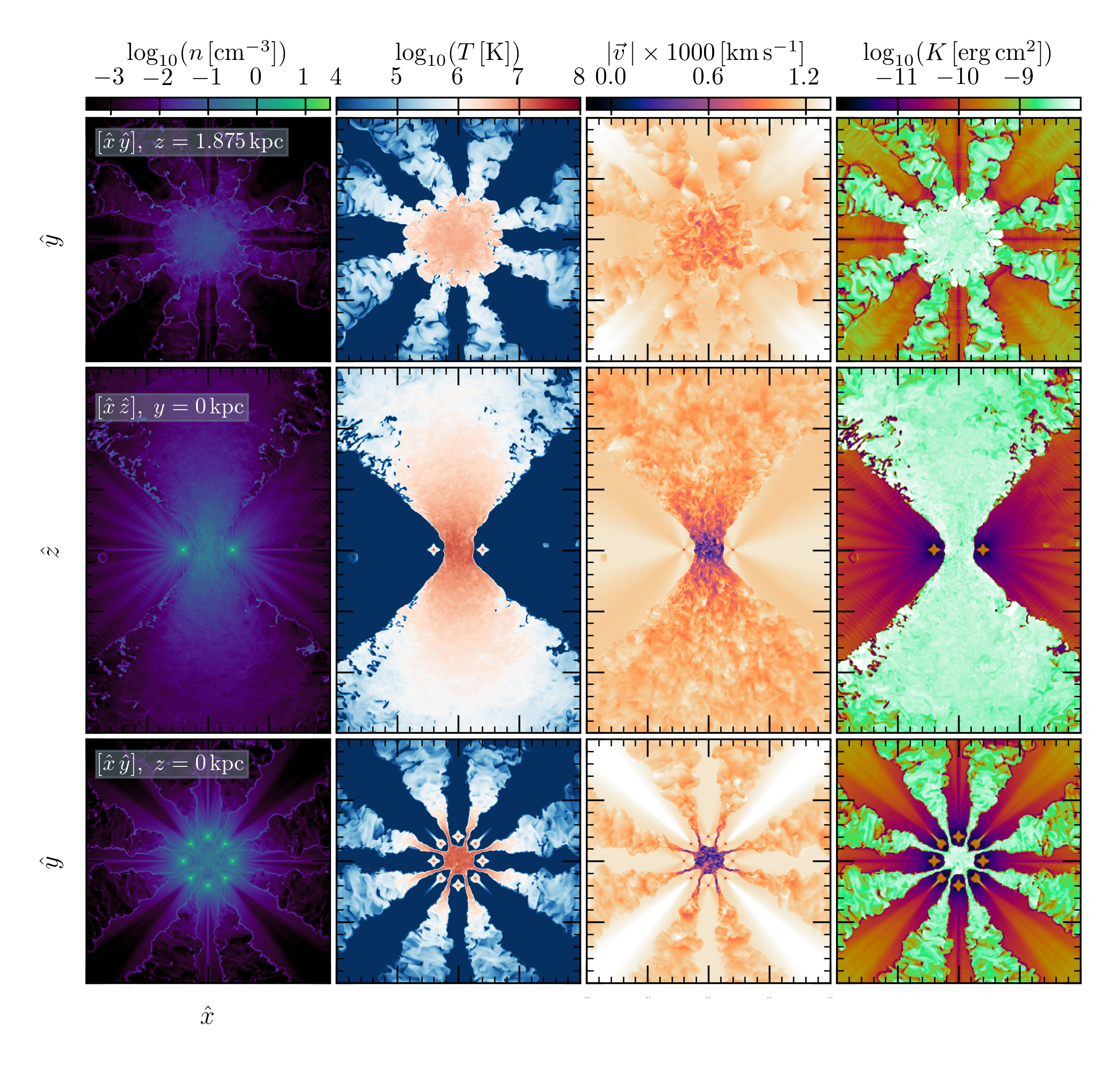}
    \caption{2\textit{d} slices of density, temperature, velocity, and entropy at 20\,Myr from Simulation `\texttt{B0}$_\star$'. The top row shows the $\hat{x}\hat{y}$ plane slice at $z=1.875\,$kpc, the middle row shows the $\hat{x}\hat{z}$ plane slice at $y=0\,$kpc, and the bottom row shows the $\hat{x}\hat{y}$ plane at $z=0\,$kpc. The ring of star clusters (bottom row) produces a multi-phase outflow (middle row) where the hot phase is enveloped with a cooler phase (top row). The simulation size is $10 \times 10 \times 15\,\mathrm{kpc}$ ($\hat{x}\times \hat{y} \times \hat{z}$). Both the top and bottom rows are $10 \times 10$\,kpc and the middle row is $10 \times 15$\,kpc.}
    \label{fig:3views_cc85a}
\end{figure}

\subsection{Non-uniform Ring}
\label{sec:nonuniform_ring}

For comparison with the uniform ring in Simulation `\texttt{A}', we contrast with Simulation `$\texttt{A}_\star$', where energy and mass is injected in 8 spherical star clusters distributed in a ring-like geometry (see Table \ref{tab:sims}). Each cluster generates a wind that is identical to that predicted by the CC85 wind model, but with strong bulk radiative cooling outside of the individual cluster volume \citep{Wang1995,Silich2004,Thompson2016}. To make a direct comparison with the uniform ring model presented in Section \ref{sec:uniform_ring}, we use the same total energy and mass-injection rates $\dot{E}_T$ and $\dot{M}_T$, appropriate for a mass-loaded SN-driven flow and a constant star formation rate. We thus neglect the time-dependent nature of cluster evolution where massive star wind precede SNe, with different relative energy and mass injection rates \citep{Silich2004,Lochhaas2018}. In Fig.~\ref{fig:3views_cc85a}, we plot three different 2\textit{d} slices of the gas number density, temperature, velocity magnitude, and entropy as in Fig.~\ref{fig:3views_plainb}. As in Simulation `\texttt{A}', after an initial transient the system reaches a quasi steady-state. The flow interaction between pairs of clusters is reminiscent of colliding winds from massive star binary star systems  \citep{Girard1987,Kenny2005}. 

The non-uniform ring develops multi-phase outflow with a starker temperature contrast than the layered wind generated by the uniform ring (in contrast with Fig.~\ref{fig:3views_plainb}). This is due to the relative mass-loading within each star-cluster. The distance at which a flow becomes radiative is critically dependent on this volumetric mass-loading rate \citep{Wang1995,Silich2004,Thompson2016}. With fixed $\dot{E}_T$ and $\dot{M}_T$, for a ring of $N$-star clusters, the cooling radius, centered on an individual star cluster, is $R_\mathrm{cool} = 81.5 \, \mathrm{pc} \, (\alpha^{2.13} / \beta^{2.92}) (N/8)^{0.79} (\mu/0.6)^{2.13} (R_A/25\,\mathrm{pc})^{1.79}$. For this simulation, we find $R_\mathrm{cool} \simeq 0.36\,$kpc. As shown in Fig.~\ref{fig:3views_cc85a}, this appears to be the case. The inner ring flow briefly undergoes rapid cooling to $10^4\,$K before it is shock heated back up to $10^7\,$K at the ring's center. Material from the cool outflow is continuously incorporated and mixed with the hot phase at the ``edges" of the cone.  Intermediate warm ($T\simeq 10^5$) temperature ``spokes" are formed by the cluster spacing. The hot flow near the top and bottom of the simulation begins to approach a cool-warm temperature, as the cooling rate becomes more dominant over the advection rate. Surprisingly, the cool phase has a higher velocity than the hot phase. Because the resolution is comparable to the radius of an individual cluster, there are grid effects. Our resolution tests suggest that the qualitative features shown here are robust to changes in resolution. Details of mixing between the hot and cooler phases may be affected by spatial resolution. 

\begin{figure}[h!]
    \centering
    \includegraphics[width=\textwidth]{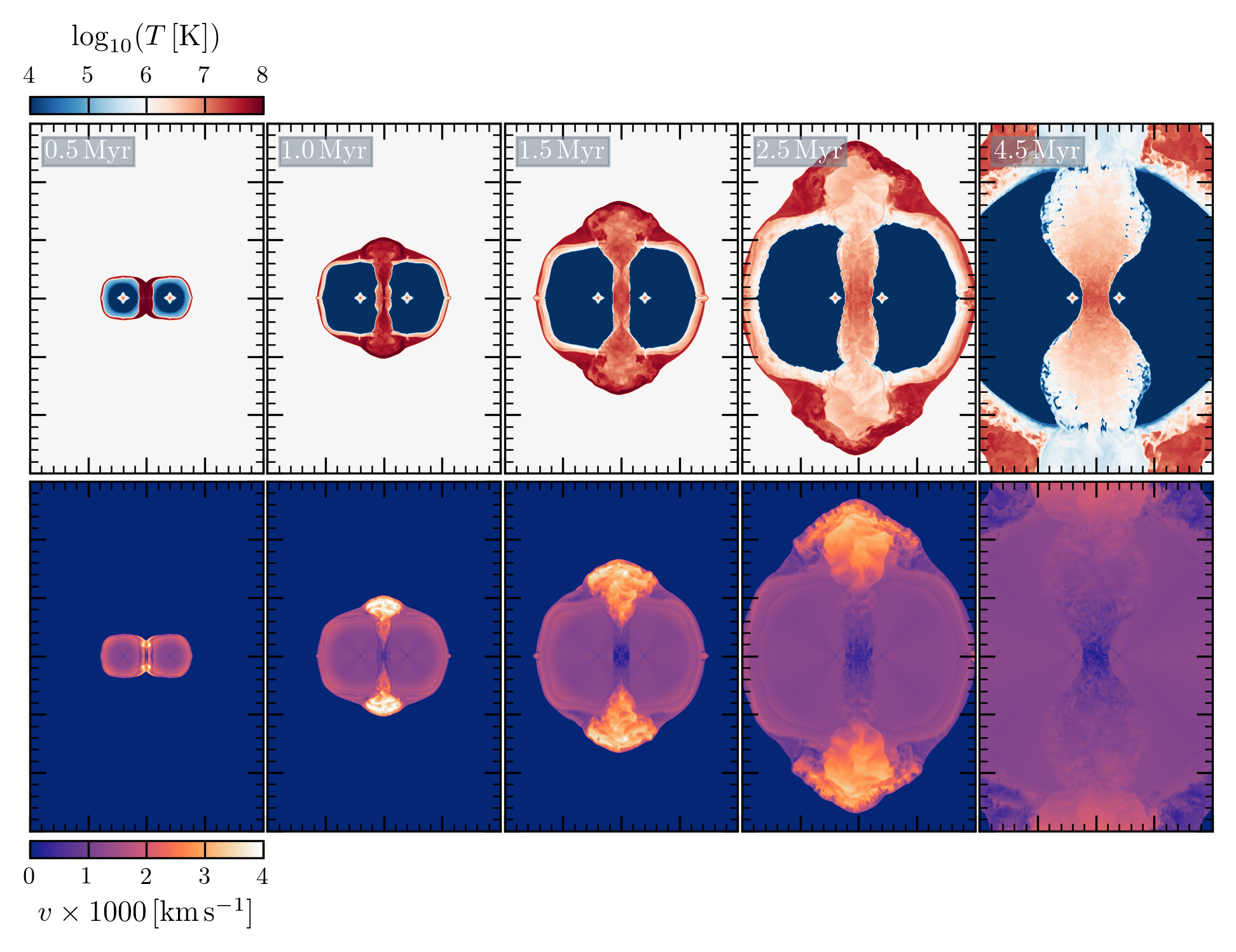}
    \caption{Temperature and velocity slices for $t=0.5, \, 1.0, \, 1.5, \, 2.5 , \, \mathrm{and} \, 4.5\,$Myr, for Simulations `$\texttt{A}_\star$'. Before the wind reaches its final time-steady form, it has a bubble-like morphology. The emergent bubble from the ring hole is fast, with peak velocities of $4000\,\mathrm{km\,s^{-1}}$, compared to the $2000\,\mathrm{km\,s^{-1}}$ shock front from the unobstructed, outer-ring, flow. The asymptotic velocity of the steady-state configuration is $v_\infty \simeq 1300 \, \mathrm{km\,s^{-1}}$. The high-velocity feature depends on the density of the surrounding medium, and the peak velocities are slower for a denser medium (see Appendix ~\ref{sec:beta_trans}). The dynamics appear consistent with expectations of a bubble driven by constant energy injection in a uniform density medium (see Eq.~\ref{eq:blast_velocity}). We see that by 4.5\,Myr, the leading bubble becomes radiative and cools faster than the shock front. This cooling fast bubble would be brighter than the surrounding shock in X-ray. The initial background at $t=0$, consists of a galactic halo-like hot gaseous background that gets blown out of the simulation domain within $\simeq$ 8\,Myr. Each panel is a slice through the $\hat{x}\hat{z}$ plane and is $10 \times 15$\,kpc.}
    \label{fig:transient}
\end{figure}

In Fig.~\ref{fig:transient}, we plot temperature and velocity slices for five different early times, $t=0.5, \, 1.0, \, 1.5, \, 2.5, \, \mathrm{and} \, 4.5\,$Myr, for Simulation `$\texttt{A}_\star$'. The simulation domain at $t=0$ is filled with an initial background of a warm, low density, gas ($n=10^{-5}\,\mathrm{cm^{-3}}$ and $T=10^6\,$K) which gets blown out of the simulation domain within $10\,$Myr. During the initial blow out, the maximum speed of the material on the computational grid is $\sim 4000\,\mathrm{km\,s^{-1}}$ at the leading edge of the collimated shock (see $t=1.0\,\mathrm{Myr}$ in Fig.~\ref{fig:transient}), which is approximately three times the asymptotic velocity of the flow driven from the individual clusters ($v\simeq 1300 \, \mathrm{km\,s^{-1}}$) or two times the non-collimated shock front ($\sim 2000\,\mathrm{km\,s^{-1}}$). These speeds occur only during the initial transient, which takes a bubble-like shape. This is not to be confused with the wind behind the bubble transient and shock front, which also appears bubble-like in the right most panel of Fig.~\ref{fig:transient}. This structure reaches its final steady-state shape within an additional Myr and will then appear like the outflows shown in  Figs.~\ref{fig:3views_plainb} and \ref{fig:3views_cc85a}. The high velocity bubble transient appears to be driven from the initial compression of the gas within the ring hole, before a steady-state rarefied volume develops. The subsequent dynamics is consistent with the expectations for a bubble driven with constant energy injection in a uniform density medium (see Appendix \ref{sec:beta_trans}). Shown in the middle panels of Fig.~\ref{fig:transient}, the deceleration of the bubble is well described by Eq.~\ref{eq:blast_velocity}, where at 1\,Myr, it is predicted that $v \sim 4300 \, \mathrm{km \, s^{-1}}$ which slows down to $v \sim 3000 \, \mathrm{km \, s^{-1}}$ at $2.5\,$Myr. The high velocity bubble leaves the simulation domain ($z=7.5\,$kpc) within $t\sim 5\,$Myr. 

\begin{figure}[h!]
    \centering
    \hspace{-1.85cm}
    \includegraphics[width=0.75\textwidth]{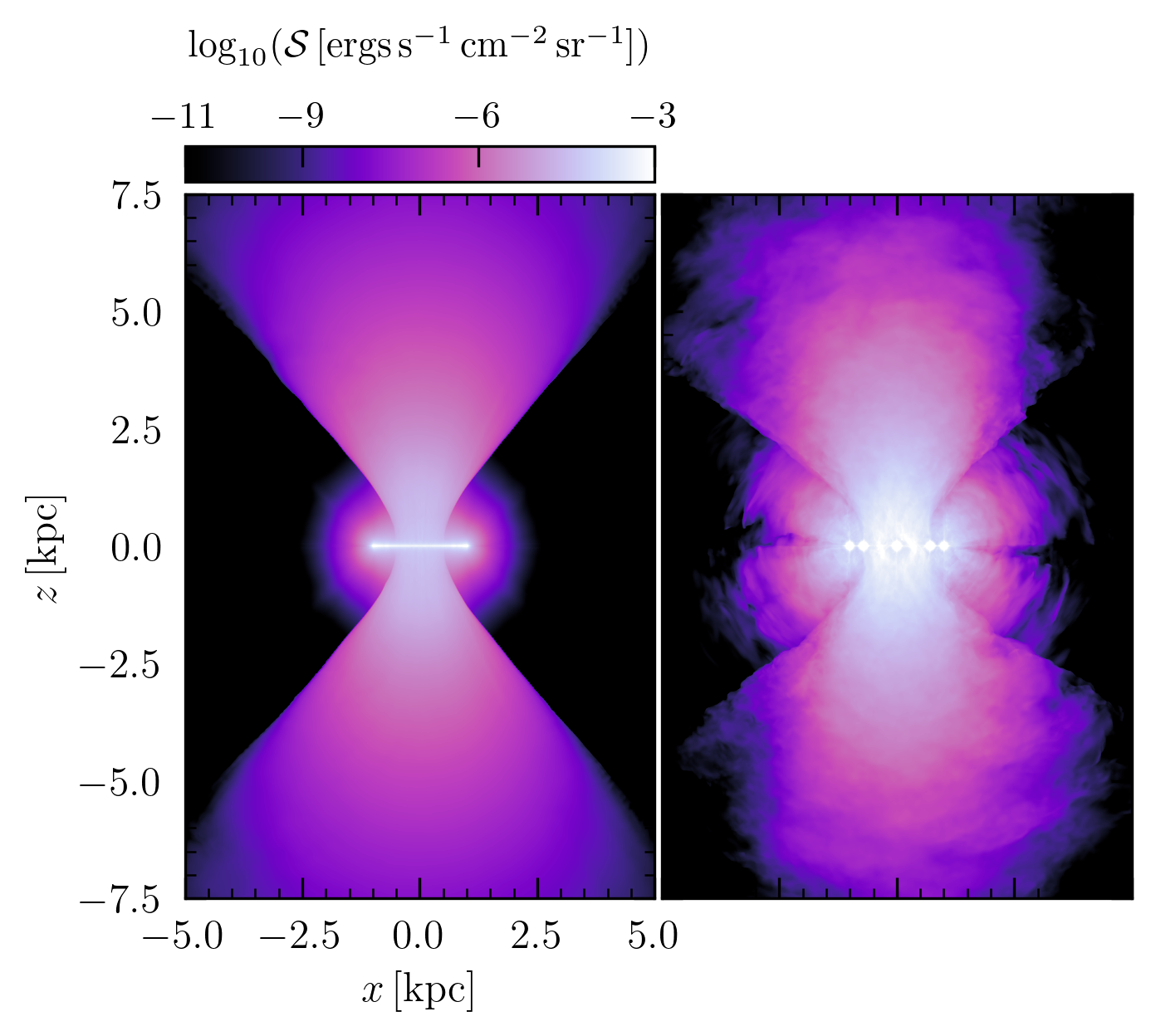}
    \caption{The X-ray surface brightness projected along the $\hat{y}$ direction which represents an ideal edge-on view of the ring system at 20$\,$Myr for Simulations `\texttt{A}' and 
    `\texttt{A}$_\star$' (from Figures \ref{fig:3views_plainb} and \ref{fig:3views_cc85a}). The simulation size is $10 \times 10 \times 15\,\mathrm{kpc}$ ($\hat{x}\times \hat{y} \times \hat{z}$).}
    \label{fig:xraybright}
\end{figure} 

In Fig.~\ref{fig:xraybright}, we calculate the instantaneous X-ray surface brightness of the steady-state winds of Simulations `\texttt{A}' and `\texttt{A}$_\star$' as $\mathcal{S}_X^{\nu_1,\nu_2}(x,z) = \int_{\nu_1}^{\nu_2}d\nu \, \int_0^{L_y}dy \, n(x,y,z)^2 \Lambda(T(x,y,z),\nu)$, where $L_y$ is the length of the simulation domain. This calculation assumes no obscuration, collisional equilibrium ionization of the plasma, solar metallicity, and optically thin-cooling. The energy band of the calculation is carried out in $E_\nu =0.3-2\,$keV. In both models, the edge-on projected surface brightness peaks at the wind driving rings, and is bright along the collimated hot outflows.

\section{Summary}

In this work we study the outflows from a starburst ring. We explore parameterized models with both uniform rings and star-clusters distributed as a ring (Section \ref{sec:results}). We consider a range of different ring geometries (Appendix \ref{sec:donut_geo}), different mass-loading rates (Appendix \ref{sec:betas}), and different background densities (Appendix \ref{sec:beta_trans}). In analogy with previous calculations of spherical outflows, unobstructed flows that are sufficiently mass-loaded (see Fig.~\ref{fig:all_betas} in Appendix \ref{sec:betas}), and/or have a sufficiently thin ring radius $R_A$ (see Fig.~\ref{fig:donut_geos} in Appendix \ref{sec:donut_geo}),  undergo rapid bulk cooling \citep{Wang1995,Silich2004,Thompson2016}. Critically, the ring model proposed here creates a layered, multi-temperature, outflow (see Figs.~\ref{fig:3views_plainb} and \ref{fig:3views_cc85a}) in which the hot ($T\geq10^7\,$K) phase is collimated and surrounded by the cool ($T=10^4\,$K) phase. This layered morphology is qualitatively similar to the that observed in nearby starbursts such as M82 \citep{Leroy2015}, NGC253 \citep{Bolatto2013}, and NGC 3079 \citep{Hodges-Kluck2020}. This model also produces a fast ($v \geq 1000\,\mathrm{km\,s^{-1}}$) cool phase.

The Milky Way is observed to have $\gamma$-ray \citep{Su2010} and X-ray \citep{Predehl2020} emitting bubbles extending several kpc above and below the disk. While the origin of the bubbles is under debate, a popular model is AGN feedback. \citet{Yang2022} showed that a single event of jet activity from the central super-massive black hole can explain the bubbles. In Fig.~\ref{fig:transient}, we show that the steady-state ring-driven wind is preceded by a transient structure that exhibits a bubble-like morphology. Here we emphasize that star-forming rings may be able to reproduce the bubble morphology, however we do not carry out a study with CMZ-like parameters and conditions for a detailed comparison. The kinematics and structure of the bubbles depend on the density of the ambient medium (see Appendix \ref{sec:beta_trans}). For background densities of $n_\mathrm{bkgrd} = 10^{-3}$ and $10^{-5}$\,cm$^{-3}$ (see Appendix \ref{sec:beta_trans}), we find that the bubbles reach velocities of $v \simeq 3,000-10,000 \, \mathrm{km\,s^{-1}}$. These velocities are within the regime of AGN-driven outflows (e.g., \citealt{King2015}) and the fast outflows seen from compact post-starburst galaxies \citep{Diamond-Stanic2012}. The dynamics of the bubbles appears to be consistent with expectations for a bubble driven by constant energy injection (see Appendix~\ref{sec:beta_trans} and Eq.~\ref{eq:blast_velocity}).

The two models highlighted in Figs.~\ref{fig:3views_plainb} and \ref{fig:3views_cc85a} employ a rather large ring width of $R_A=25\,$pc. The radius of a typical super-star cluster in M82 is much smaller, at around $5\,$pc \citep{Smith2006}. A limitation of our work is the spatial resolution ($\Delta x = 13.0\,$pc). As a result, we cannot accurately model individual realistic clusters with much smaller radii. This limitation is apparent in Simulations `\texttt{A}' and `$\texttt{A}_\star$', where grid effects are present (see Figs.~\ref{fig:3views_plainb} and \ref{fig:3views_cc85a}). \citet{Silich2004} found that radiative outflows should be generic for super star clusters with mass and energy injection by stellar winds. We note that we do not model the time-dependent evolution of clusters, where a wind phase precedes SN energy and mass injection. To make the comparison between models `\texttt{A}' and `$\texttt{A}_\star$' most clear, we treat the properties of the energy and mass injection as constant over the short time required to establish a steady-state structure. A future study will include different evolutionary phases of each star cluster (as in \citet{Schneider2020}). 


Our simulations are simple parameterized models that neglect some pieces of physics in an effort to sketch the main components of winds driven by star-forming rings, and to contrast with the spherical CC85 model. Probably the most important for the qualitative character of the outflow is that we neglect the gas disk that would surround any star-forming ring. The gas densities of a typical disk are several orders magnitude larger than that of the initialized background in our simulations (see Section \ref{sec:beta_trans}), and may seed the hot and cool outflows with clumpy structures provided by the wind-ISM interaction \citep{Cooper2007,Tanner2016,Tanner2022}. A future work will explore simulations including both a disk and more realistic ISM and kpc-scale CGM structures. Because of the large velocities obtained in these simulations, we have also neglected gravity. For models with lower thermalization efficiency and higher mass-loading, the outflows will have lower velocities for which gravity may become important. We have further used a simple approximation radiative cooling, and have neglected non-equilibrium ionization and irradiation effects, which is important to the ionization state of the outflow \citep{Gray2019,Sarkar2022}.

\begin{acknowledgments}

DDN and TAT thank the OSU Galaxy/ISM Meeting for useful discussions. DDN thanks Evan Schneider for guidance with running \texttt{Cholla} simulations. We thank Ryan Tanner and Rebecca Levy for insightful conversations. We thank the anonymous referee for comments that have improved the manuscript. DDN thank Ohio Supercomputing Center engineer, Zhi-Qiang Yuo, for technical assistance. DDN and TAT are supported by National Science Foundation Grant \#1516967, NASA ATP 80NSSC18K0526, and NASA 21-ASTRO21-0174.
\end{acknowledgments}

\vspace{5mm}

\software{\texttt{numpy} \citep{Harris2020numpy}, 
          \texttt{scipy} \citep{2020SciPy-NMeth} ,  
          \texttt{Cholla} \citep{Schneider2015},
          \texttt{pyatomdb} \citep{2020pyatomdb} , 
          \texttt{yt} \citep{2011Turk_yt}
          } 
\newpage

\appendix

\section{The uniform starburst ring geometry and volumetric injection rates}
\label{sec:EdotMdot}

The physical system is specified by energy and mass-injection within a ring. We choose to specify the starburst volume in a method similar to that in CC85. In the spherical CC85 model, the problem can be viewed in one-dimension, namely, the region of energy and mass-injection is satisfied by the bounds: $r \leq R$, where $R$ is the radius of the wind-driving sphere \citep{Chevalier1985}. The ring requires additional variables. The controlling parameters of the ring system are the distance from the origin to the center of the ring, $R_B$, and the radius of the ring, $R_A$. For a given $R_A$ and $R_B$, we determine the minimum and maximum radii, $R_\mathrm{min} (\Gamma)$ and $R_\mathrm{max} (\Gamma)$, which bisects outer and inner ring surfaces, respectively. For specific line of sight angle, $\Gamma$, the line of sight chord endpoints, which skewers the ring surface twice, are derived to be $R_\mathrm{min} = R_\mathrm{B} \cos (\Gamma) - (R_\mathrm{A}^2 - R_\mathrm{B}^2/2 + R_\mathrm{B}^2 \cos (2 \Gamma) / 2 )^{1/2}$ and $R_\mathrm{max} = R_\mathrm{B} \cos (\Gamma) + (R_\mathrm{A}^2 - R_\mathrm{B}^2/2 + R_\mathrm{B}^2 \cos (2 \Gamma)/ 2 )^{1/2}.$ The bounds, in spherical coordinates, corresponding to the ring volume are derived to be $R_\mathrm{min} \leq r \leq R_\mathrm{max}, \ 0 \leq \theta \leq \pi, \ 0 \leq \varphi \leq 2 \pi,$ where $ 0 \leq \Gamma \leq \Gamma_\mathrm{crit}, \ \Gamma = \pi/2 - \theta, \ \mathrm{and} \ \Gamma_\mathrm{crit} = \tan^{-1} ( R_A /R_B ).$ This is done for every sight line $\Gamma (\theta)$ and then revolved over the azimuthal axis. We see that for opening angle $\Gamma = 0$, we recover the expectation that $R_\mathrm{min} = R_B - R_A$ and $R_\mathrm{max} = R_B +R_A$. The uniform volumetric energy and mass-injection rates are then:
\begin{align}
    & q(r,\theta) = \begin{cases}
    \dot{M}_\mathrm{hot}/ V,  & R_\mathrm{min} \leq r \leq R_\mathrm{max} \\
    0, & R_\mathrm{max} < r \ \mathrm{or} \ r < R_\mathrm{min}
    \end{cases} \label{eq:qdot} \\ 
    & Q(r,\theta) = \begin{cases}
    \dot{E}_\mathrm{hot}/ V,  & R_\mathrm{min} \leq r \leq R_\mathrm{max} \\
    0, & R_\mathrm{max} < r \ \mathrm{or} \ r < R_\mathrm{min} 
    \end{cases}  \label{eq:Qdot}
\end{align}
where the normalizations for $\dot{E}_T$ and $\dot{M}_T$ are described in Section \ref{sec:methods}, and $V$ is the volume of the starburst ring:
\begin{equation}
    V = 2 \pi ^2 R_A^2 R_B.
\end{equation}

\section{Varied Ring Geometries}
\label{sec:donut_geo}
\subsection{Varied Uniform Ring Geometries} 
\begin{figure}[h!]
    \centering
    \includegraphics[width=\textwidth]{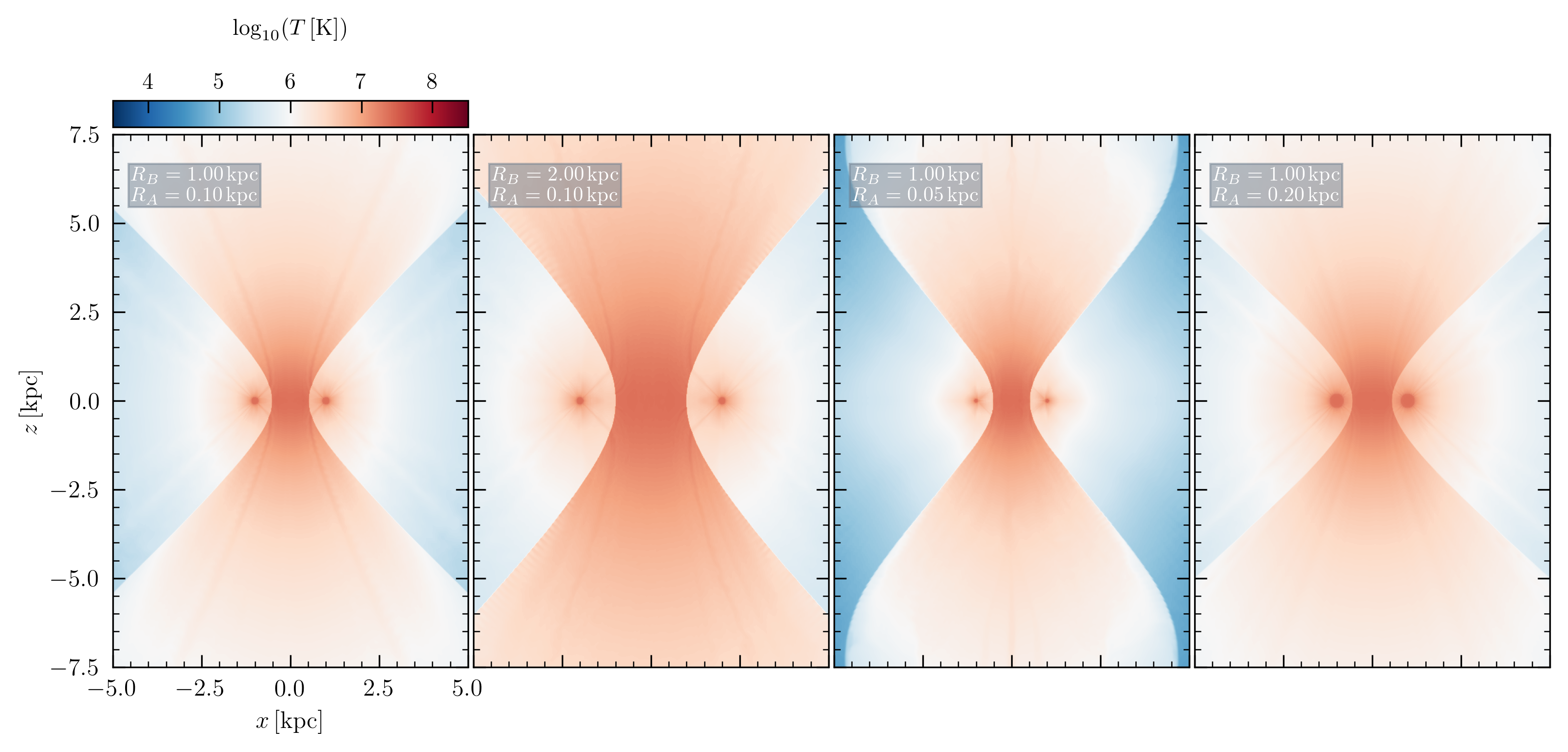}
    \caption{2\textit{d} temperature slices of Simulations `\texttt{B0}', `\texttt{B2}', `\texttt{B3}', and `\texttt{B4}'. We see that the different ring geometries produce different outflow morphologies despite equivalent energy and mass-injection rates $\alpha$ and $\beta$.}
    \label{fig:donut_geos}
\end{figure}

In Fig.~\ref{fig:donut_geos}, we show 2\textit{d} temperature slices for Simulations `\texttt{B0}', `\texttt{B2}', `\texttt{B3}', and `\texttt{B4}' (see Table \ref{tab:sims}). As we would expect, changing the ring hole radius $R_B$ affects the width of the cylindrical hot outflow. Smaller $R_A$ values lead to a smaller cooling radius, thus a more prominent temperature contrast between the unobstructed wind and the hot thermalized wind along the minor axis. Simulation `\texttt{B1}' (not plotted) similarly produces an expected morphology. The most radiative model, Simulation `\texttt{B3}' produces an hour-glass like morphology.

\section{Varied Mass-loading Rates} 
\label{sec:betas}
\begin{figure*}[h!]
    \centering
    \hspace{-1.25cm}
    \includegraphics[width=1.0\textwidth]{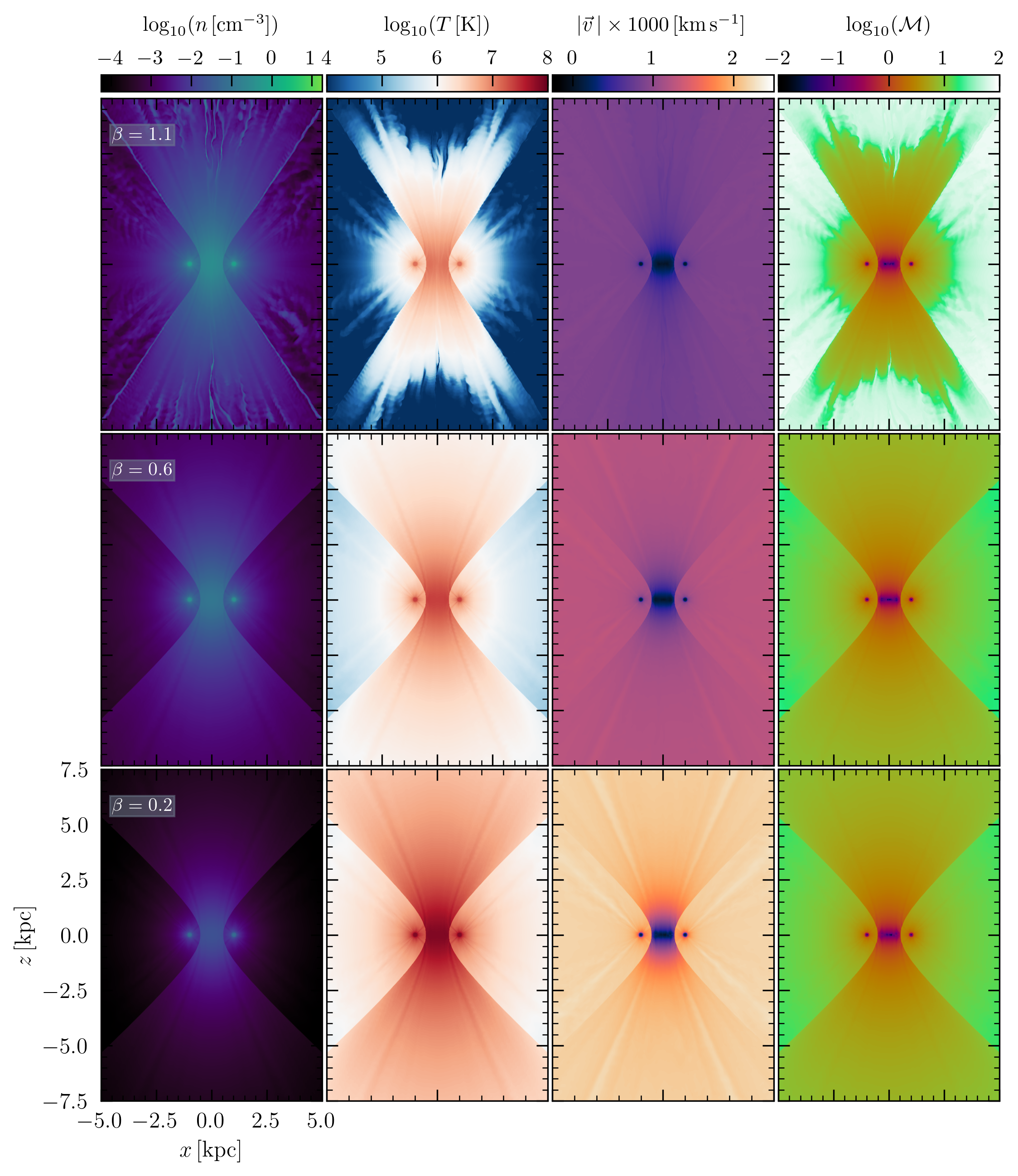}    \caption{2\textit{d} slices of density, temperature, velocity, and Mach number at 40\,Myr for varied energy thermalization efficiencies within the starburst ring: $\beta = 1.1, \ 0.6, \ \mathrm{and} \ 0.2$ (Simulations `\texttt{B5}', `\texttt{B0}', and `\texttt{B6}' respectively, see Table \ref{tab:sims}). All models have the same ring geometry of $R_A=0.025\,\mathrm{kpc}$ and $R_C=1.0\,\mathrm{kpc}$.}    \label{fig:all_betas}
\end{figure*}

In Fig.~\ref{fig:all_betas}, we show 2\textit{d} density, temperature, velocity, and Mach number slices of Simulations `\texttt{B5}', `\texttt{B0}', and `\texttt{B6}'. As detailed in Table \ref{tab:sims}, these simulations have varied the SN mass-loading rates described by $\beta=1.1, \ 0.6, \ \mathrm{and} \ 0.2$, respectively. The ring geometry produces the layered wind morphology with a temperature contrast dependent on the mass-loading. Higher mass-loading leads to a stronger contrast. The non-radiative ring wind models to quickly approach a steady-state. The highly mass-loaded wind model approaches a quasi-steady state, with small thermal instability filaments forming within the bi-cone. As shown in Fig.~\ref{fig:all_betas}, different mass-loading leads to different outflow morphologies. We fit the flared-cylinder function $A(z) = A_0 [ 1+ (z/z_b)]^\Delta$ to the simulations. For $\beta=1.1$, we find $A_0=0.56\,\mathrm{kpc^2}$, $z_b=0.51\,$kpc, and $\Delta=1.8$. For both $\beta=0.2$ and $\beta=0.6$, we find $A_0=0.85\,\mathrm{kpc^2}$, $z_b=0.75\,$kpc, and $\Delta=2.3$. We find the highest mass-loaded model $\beta=1.1$ produces the narrowest cylindrical hot outflow along the ring axis. Both the $\beta=0.2$ and $\beta=0.6$ produce equivalently-sized hot cylindrical outflows. Highly radiative models appear to produce narrower collimated hot outflows than non-radiative models.

\section{Varied initialized ambient backgrounds} 
\label{sec:beta_trans} 
Here, we vary the initialized background gas density to probe the initial interaction between the flow and ambient background during the transient phase, before the steady-state wind is established, to understand how fast velocities at the leading edge along the minor axis depend on the density of the surrounding medium. As detailed in Table \ref{tab:sims}, all simulations here have the same energy and mass-injection rates and equivalent ring geometries. Simulations `\texttt{C0}', `\texttt{C1}', and `\texttt{C2}' are initialized with constant background density of $n_\mathrm{bkgd.}=10^{-3}, \ 10^{-5}, \ \mathrm{and} \ 10^{-7} \, \mathrm{cm^{-3}}$. The temperature of the background is $T_\mathrm{bkgd.}=10^6\,$K. For background densities of $n_\mathrm{bkgd.}=10^{-3}, \ 10^{-5}, \ \mathrm{and} \ 10^{-7} \, \mathrm{cm^{-3}}$, the respectively maximum velocity of the leading bubble edge along the minor axis of the ring is approximately $5,000, \ 10,000, \ \mathrm{and} \ 15,000\,\mathrm{km\,s^{-1}}$. The physical size of the bubble $R$ at a given time after the start of the simulation is also inversely related to the density of the surrounding medium. In addition, as in the middle three panels of Fig.~ \ref{fig:transient}, the velocity of the fastest material decreases with time as the bubble edge expands. All of these features of the high-velocity transient are consistent with a bubble driven by a constant energy injection rate $\dot{E}_T$, where we expect the radius of a spherical bubble to scale as \citep{Weaver1977}
\begin{equation}
   R\sim(\dot{E}_T t^3/\rho)^{1/5}\simeq7\,{\rm kpc}\,(\dot{E}_T/3\times10^{42}\,{\rm ergs/s})^{1/5}\,(t/{\rm Myr})^{3/5}\,(10^{-5}\,{\rm cm^{-3}}/n)^{1/5},
\label{eq:blast_radius}
\end{equation}
and the velocity of the bubble edge to scale as 
\begin{equation}
dR/dt\sim \frac{3}{5} R/t\sim \frac{3}{5}(\dot{E}_T/t^2/\rho)^{1/5}\simeq4300\,{\rm km/s}\,(\dot{E}_T/3\times10^{42}\,{\rm ergs/s})^{1/5}\,({\rm Myr}/t)^{2/5}\,(10^{-5}\,{\rm cm^{-3}}/n)^{1/5}.
\label{eq:blast_velocity}
\end{equation}
Expansion through a denser medium leads to slower bubble velocities, smaller bubbles, and deceleration during expansion. The asymmetric ring geometry and the associated asymmetric velocity structure seen in Figure \ref{fig:transient} between the ring's minor and major axes complicates a more careful comparison with analytic expectations.

\newpage

\bibliography{sample631}{}
\bibliographystyle{aasjournal}

\end{document}